\documentclass[journal,a4paper,oneside,onecolumn]{IEEEtran} 
\usepackage{amsmath,amsfonts,amssymb}
\usepackage{algorithm}
\usepackage{algpseudocode}
\makeatletter
\let\OldStatex\Statex
\renewcommand{\Statex}[1][3]{%
 \setlength\@tempdima{\algorithmicindent}%
 \OldStatex\hskip\dimexpr#1\@tempdima\relax}
\makeatother
\algnewcommand{\IIf}[1]{\State\algorithmicif\ #1\ \algorithmicthen}
\algnewcommand{\EndIIf}{\unskip\ \algorithmicend\ \algorithmicif}
\algnewcommand{\FFor}[1]{\State\algorithmicfor\ #1\ \algorithmicdo}
\algnewcommand{\EndFFor}{\unskip\ \algorithmicend\ \algorithmicfor}
\algnewcommand{\WWhile}[1]{\State\algorithmicwhile\ #1\ \algorithmicdo}
\algnewcommand{\EndWWhile}{\unskip\ \algorithmicend\ \algorithmicwhile}

\newtheorem{thm}{Theorem}
\newtheorem{lem}{Lemma}
\newtheorem{rem}{Remark}

\newcommand{\lrb}[1]{\left\{{#1}\right\}}
\newcommand{\lrsb}[1]{\left({#1}\right)}

\newcommand{\Prob}{\mathrm{Prob}}
\newcommand{\I}{\mathcal{I}}
\newcommand{\Y}{\mathcal{Y}}

\newcommand{\ff}{\boldsymbol{f}}
\newcommand{\bs}{\boldsymbol{s}}
\newcommand{\uu}{\boldsymbol{u}}
\newcommand{\UU}{\boldsymbol{U}}
\newcommand{\xx}{\boldsymbol{x}}
\newcommand{\XX}{\boldsymbol{X}}
\newcommand{\yy}{\boldsymbol{y}}
\newcommand{\YY}{\boldsymbol{Y}}

\newcommand{\ttb}{\mathtt{b}}
\newcommand{\ttc}{\mathtt{c}}
\newcommand{\tti}{\mathtt{i}}
\newcommand{\ttk}{\mathtt{k}}
\newcommand{\ttM}{\mathtt{M}}
\newcommand{\ttp}{\mathtt{p}}
\newcommand{\ttP}{\mathtt{P}}
\newcommand{\ttu}{\mathtt{u}}
\newcommand{\ttU}{\mathtt{U}}
\newcommand{\ttV}{\mathtt{V}}
\newcommand{\ttX}{\mathtt{X}}

\newcommand{\ttT}{\mathtt{\Theta}}
\newcommand{\ttL}{\mathtt{\Lambda}}
\newcommand{\ttl}{\mathtt{\lambda}}

\newcommand{\updateU}{\mathtt{updateU}}
\newcommand{\updateT}{\mathtt{update\Theta}}
\newcommand{\copyPath}{\mathtt{copyPath}}
\newcommand{\extendPath}{\mathtt{extendPath}}
\newcommand{\splitPath}{\mathtt{splitPath}}
\newcommand{\prunePath}{\mathtt{prunePath}}
\newcommand{\markPath}{\mathtt{markPath}}
\newcommand{\selectPath}{\mathtt{selectPath}}
\newcommand{\calcV}{\mathtt{calcV}}
\newcommand{\Index}{\mathtt{Index}}
\newcommand{\Active}{\mathtt{Active}}
\newcommand{\magnifyP}{\mathtt{magnifyP}}
\newcommand{\ttleft}{\mathtt{left}}
\newcommand{\ttright}{\mathtt{right}}
\newcommand{\maxP}{\mathtt{maxP}}
\newcommand{\partition}{\mathtt{partition}}
\newcommand{\swapIndex}{\mathtt{swapIndex}}
\newcommand{\parity}{\mathtt{parity}}
\newcommand{\lsb}{\mathtt{lsb}}
\newcommand{\br}{\mathtt{br}}

\newcommand{\hb}{\widehat{b}}
\newcommand{\hu}{\widehat{u}}
\newcommand{\huu}{\widehat{\uu}}
\newcommand{\hx}{\widehat{x}}
\newcommand{\hxx}{\widehat{\xx}}

\allowdisplaybreaks

\begin{document}

\title{
 Successive-Cancellation Decoding\\
 of Binary Polar Codes\\
 Based on Symmetric Parametrization
}
\author{
 Jun~Muramatsu~\IEEEmembership{Senior Member,~IEEE}
 \thanks{J.~Muramatsu is with
  NTT Communication Science Laboratories, NTT Corporation,
  2-4, Hikaridai, Seika-cho, Soraku-gun, Kyoto, 619-0237 Japan
  (E-mail: jun.muramatsu@ieee.org).
}}

\maketitle

\begin{abstract}
 This paper introduces algorithms for the successive-cancellation
 decoding and the successive-cancellation list decoding
 of binary polar source/channel codes.
 By using the symmetric parametrization of conditional probability,
 we reduce both space and time complexity
 compared to the original algorithm introduced by Tal and Vardy.
\end{abstract}

\section{Introduction}
Polar source/channel codes were introduced by Ar\i{}kan~\cite{A09,A10,A11}.
When these codes are applied to source coding with decoder side information
for joint stationary memoryless sources,
the coding rate achieves a fundamental limit called the conditional entropy.
When applied to the channel coding of a symmetric channel,
the coding rate achieves a fundamental limit called the channel capacity.
Ar\i{}kan introduced successive-cancellation decoding,
which can be implemented with computational
complexity of $O(N\log_2 N)$ where $N$ is block length.

In this paper, we introduce algorithms for successive-cancellation
decoding and successive-cancellation list decoding
based on the work of Tal and Vardy \cite{TV15}.
Our constructions can be applied to both polar source codes
and polar channel codes.
Furthermore, the proposed list decoding algorithm
reduces the space and time complexity compared to \cite{TV15}.

\section{Definitions and Notations}
Throughout this paper, we use the following definitions and notations.

For a given $n$, let $N\equiv 2^n$ denote the block length.
We assume that the number $n$ is given as a constant,
which means that all algorithms have access to this number.
We use the bit-indexing approach introduced in \cite{A09}.
The indexes of a $N$-dimensional vector are represented by
$n$-bit sequences as $X^N\equiv(X_{0^n},\ldots,X_{1^n})$,
where $0^n$/$1^n$ denotes the $n$-bit all zero/one sequence.
To represent an interval of integers, we use the following notations
\begin{align*}
 [0^n:b^n]
 &\equiv
 \{0^n,\ldots,b^n\}
 \\
 [0^n:b^n)
 &\equiv
 [0^n:b^n]\setminus\{b^n\}.
 \\
 [b^n:1^n]
 &\equiv
 [0^n:1^n]\setminus[0^n:b^n)
 \\
 (b^n:1^n]
 &\equiv
 [0^n:1^n]\setminus[0^n:b^n].
\end{align*}
For a given subset $\I$ of $[0^n:1^n]$,
we define the sub-sequences of $X^N$ as
\begin{equation*}
 X_{\I}
 \equiv\{X_{b^n}\}_{b^n\in\I}.
\end{equation*}
Let $c^lb^k\in\{0,1\}^{l+k}$ be the concatenation of 
$b^k\in\{0,1\}^k$ and $c^l\in\{0,1\}^l$.
For given $b^k\in\{0,1\}^k$ and $c^l\in\{0,1\}^l$,
we define subsets $c^l[0:b^k]$ and $c^l[0:b^k)$ of $\{0,1\}^{k+l}$ as
\begin{align*}
 c^l[0:b^k]
 &\equiv
 \{c^ld^k: d^k\in[0^k:b^k]\}
 \\
 c^l[0:b^k)
 &\equiv
 \{c^ld^k: d^k\in[0^k:b^k)\}.
\end{align*}
The bipolar-binary conversion, $\mp_b$, of $b\in\{0,1\}$ is defined as
\begin{align*}
 \mp_b
 &\equiv
 \begin{cases}
  -
  &\text{if $b=1$}
  \\
  +
  &\text{if $b=0$}.
 \end{cases}
\end{align*}

\section{Binary Polar Codes}
\label{sec:polar-code}
In this section, we revisit the binary polar source/channel codes introduced
in previous works~\cite{A09,A10,A11,S12}.

Assume that $\{0,1\}$ is the binary finite field.
For given positive integer $n$, polar transform $G$ is defined as
\begin{equation*}
 G\equiv
 \begin{pmatrix}
  1 & 0
  \\
  1 & 1
 \end{pmatrix}^{\otimes n}
 \Pi_{\mathrm{BR}},
\end{equation*}
where $\otimes n$ denotes the $n$-th Kronecker power
and $\Pi_{\mathrm{BR}}$ is the bit-reversal permutation matrix~\cite{A09}.
Next, vector $\uu\in\{0,1\}^N$ is defined as $\uu\equiv \xx G$
for given vector $\xx\in\{0,1\}^N$.
For completeness, an algorithm that computes $\uu$
is given in Appendix~\ref{sec:polar-transform}.
Let $\{\I_0,\I_1\}$ be a partition of $[0^n:1^n]$,
satisfying $\I_0\cap\I_1=\emptyset$ and $\I_0\cup\I_1=[0^n:1^n]$.
We define $\{\I_0,\I_1\}$ later.

Let $\XX\equiv(X_{0^n},\ldots,X_{1^n})$ and $\YY\equiv(Y_{0^n},\ldots,Y_{1^n})$
be random variables and let $\UU\equiv(U_{0^n},\ldots,U_{1^n})$
be a random variable defined as $\UU\equiv \XX G$.
Then $P_{U_{\I_0}U_{\I_1}\YY}$,
the joint distribution of $(U_{\I_0},U_{\I_1},\YY)$,
is defined using the joint distribution $P_{\XX\YY}$ of $(\XX,\YY)$ as
\begin{equation*}
 P_{U_{\I_0}U_{\I_1}\YY}(u_{\I_0},u_{\I_1},\yy)
 \equiv P_{\XX\YY}((u_{\I_1},u_{\I_0})G^{-1},\yy),
\end{equation*}
where the elements in $(u_{\I_1},u_{\I_0})$ are sorted in
index order before operation $G^{-1}$.
We refer to $u_{\I_1}$ and $u_{\I_0}$ as
{\it frozen bits} and {\it unfrozen bits}, respectively.

Let $P_{U_{b^n}|U_{[0^n:b^n)}\YY}$ be the conditional probability distribution,
defined as
\begin{align*}
 P_{U_{b^n}|U_{[0^n:b^n)}\YY}(u_{b^n}|u_{[0^n:b^n)},\yy)
 &\equiv
 \frac{\sum_{u_{(b^n:1^n]}}P_{U_{\I_0}U_{\I_1}\YY}(u_{\I_0},u_{\I_1},\yy)}
 {\sum_{u_{[b^n:1^n]}}P_{U_{\I_0}U_{\I_1}\YY}(u_{\I_0},u_{\I_1},\yy)}.
\end{align*}
For vector $u_{\I_1}$ and side information $\yy\in\Y^N$,
output $\huu\equiv \ff(u_{\I_1},\yy)$ of
successive-cancellation (SC) decoder $\ff$ is defined recursively as
\begin{equation*}
 \hu_{b^n}
 \equiv
 \begin{cases}
  f_{b^n}(\hu_{[0^n:b^n)},\yy)
  &\text{if}\ b^n\in\I_0
  \\
  u_{b^n}
  &\text{if}\ b^n\in\I_1
 \end{cases}
\end{equation*}
using function $\{f_{b^n}\}_{b^n\in\I_0}$ defined as
\begin{equation*}
 f_{b^n}(u_{[0^n:b^n)},\yy)
 \equiv
 \arg\max_{u\in\{0,1\}}P_{U_{b^n}|U_{[0^n:b^n)}\YY}(u|u_{[0^n:b^n)},\yy),
\end{equation*}
which is the maximum a posteriori decision rule
after an observation $(u_{[0^n:b^n)},\yy)$.

For a polar source code (with decoder side information),
$\xx\in\{0,1\}^N$ is a source output, $u_{\I_1}$ is a codeword,
and $\yy\in\Y^N$ is a side information output.
The decoder reproduces source output $\hxx\equiv\ff(u_{\I_1},\yy)G^{-1}$ 
from codeword $u_{\I_1}$ and $\yy$.
The (block) decoding error probability is given as
$\Prob(\ff(U_{\I_1},\YY)G^{-1}\neq \XX)$.

For a systematic polar channel code \cite{A11},
we define $\I'_0$ and $\I'_1$ as
\begin{align}
 \I'_0
 &\equiv\lrb{
  b_0b_1\cdots b_{n-1}: b_{n-1}\cdots b_1b_0\in\I_0
 }
 \label{eq:I'0}
 \\
 \I'_1
 &\equiv\lrb{
  b_0b_1\cdots b_{n-1}: b_{n-1}\cdots b_1b_0\in\I_1
 }
 \notag
\end{align}
for given $(\I_0,\I_1)$.
We assume that encoder and decoder share a vector $u_{\I_1}$.
The encoder computes $(x_{\I'_1},u_{\I_0})$
from message $x_{\I'_0}$ and shared vector $u_{\I_1}$
so that $(x_{\I'_1},x_{\I'_0})=(u_{\I_1},u_{\I_0})G^{-1}$,
where the elements in $(x_{\I'_1},x_{\I'_0})$ and 
$(u_{\I_1},u_{\I_0})$ are sorted in index order before operating $G^{-1}$.
An algorithm for this computation is introduced
in Appendix \ref{sec:systematic}.
The encoder then generates channel input $\xx\equiv(x_{\I'_0},x_{\I'_1})$,
where the elements in $(x_{\I'_1},x_{\I'_0})$ are sorted in index order.
The decoder reproduces channel input $\hxx\equiv \ff(u_{\I_1},\yy)G^{-1}$
from channel output $\yy\in\Y^N$ and shared vector $u_{\I_1}$,
where $\hx_{\I'_0}$ is a reproduction of the message.
The (block) decoding error probability is also given as
$\Prob(\ff(U_{\I_1},\YY)G^{-1}\neq \XX)$.

For a non-systematic polar channel code,
$u_{\I_0}$ is a message
and vector $u_{\I_1}$ is shared by the encoder and decoder.
The encoder generates channel input $\xx\in\{0,1\}^N$
as $\xx\equiv(u_{\I_1},u_{\I_0})G^{-1}$,
where the elements in $(u_{\I_1},u_{\I_0})$ are sorted in
index order before operating $G^{-1}$.
The decoder reproduces the pair of vectors
$(u_{\I_1},\hu_{\I_0})\equiv \ff(u_{\I_1},\yy)$
from channel output $\yy\in\Y^N$ and shared vector $u_{\I_1}$,
where $\hu_{\I_0}$ is a reproduction of the message.
The (block) decoding error probability is given as
$\Prob\lrsb{\ff(U_{\I_1},\YY)\neq (U_{\I_0},U_{\I_1})}$.

We have the following lemmas.
\begin{lem}[{\cite[Theorem 2]{A10},\cite[Theorem 4.10]{S12}}]
Define $\I_0$ as
\begin{align*}
 \I_0
 &\equiv\lrb{
  b^n\in[0^n:1^n]:
  Z(U_{b^n}|U_{[0^n:b^n)},Y_{[0^n:1^n]})\leq 2^{-2^{n\beta}}
 },
\end{align*}
where $Z(U_{b^n}|U_{[0^n:b^n)},Y_{[0^n:1^n]})$
is the source Bhattacharyya parameter introduced in \cite{A10}.
Then we have 
\begin{align*}
 \lim_{n\to\infty}\frac{|\I_0|}{2^n}
 &=
 1-H(X|Y)
 \\
 \lim_{n\to\infty}\frac{|\I_1|}{2^n}
 &=
 H(X|Y)
\end{align*}
for any $\beta\in[0,1/2)$.
\end{lem}
\begin{lem}[{\cite[Proposition 2.7]{S12}}]
\begin{equation*}
 \Prob(f_{b^n}(U_{[0^n:b^n)},\YY)\neq U_{b^n})
 \leq
 Z(U_{b^n}|U_{[0^n:b^n)},Y_{[0^n:1^n]}).
\end{equation*}
\end{lem}

We have the following lemma, which can be shown as in 
a previous proof~\cite{A09}.
\begin{lem}[{\cite[Lemma 2]{SCD},\cite[Eq.~(1)]{TV13}}]
\begin{align*}
 \Prob(\ff(U_{\I_1},\YY)G^{-1}\neq \XX)
 &=
 \Prob\lrsb{\ff(U_{\I_1},\YY)\neq (U_{\I_0},U_{\I_1})}
 \notag
 \\
 &\leq
 \sum_{b^n\in\I_0}
 \Prob(f_{b^n}(U_{[0^n:b^n)},\YY)\neq U_{b^n}).
\end{align*}
\end{lem}

From the above lemmas,
we have the fact that the rate of polar codes attains the fundamental limit
and the decoding error probability goes to zero as $n\to\infty$.
For example, we can obtain $\I_0$ by using the technique
introduced in \cite{POLARIZATION,TV13}
In the following sections, we assume that $\I_0$ is given arbitrary.

\section{Symmetric Parametrization}

In this section, we introduce the polar transform
based on symmetric parametrization.
Given $P_U$, a probability distribution of binary random variable $U$,
let $\theta$ be defined as
\begin{equation*}
 \theta
 \equiv
 P_U(0) - P_U(1).
\end{equation*}
Then we have
\begin{equation*}
 P_U(u)
 =
 \frac{1\mp_u\theta}2,
\end{equation*}
where $\mp_u$ is the bipolar-binary conversion of $u$.

In the basic polar transform,
a pair of binary random variables $(U_0,U_1)$ is transformed into
\begin{align*}
 U'_0
 &\equiv
 U_0\oplus U_1
 \\
 U'_1
 &\equiv
 U_1,
\end{align*}
where $\oplus$ denotes the addition on the binary finite field.
Assume that random variables $U_0,U_1\in\{0,1\}$ are independent.
For each $i\in\{0,1\}$, let $\theta_i$ be defined as
\begin{equation*}
 \theta_i
 \equiv
 P_{U_i}(0) - P_{U_i}(1).
\end{equation*}
First, we have
\begin{align}
 P_{U'_0}(0)
 &=
 P_{U_0}(0)P_{U_1}(0) + P_{U_0}(1)P_{U_1}(1)
 \notag
 \\
 &=
 \frac{1+\theta_0}2\cdot\frac{1+\theta_1}2
 + \frac{1-\theta_0}2\cdot\frac{1-\theta_1}2
 \notag
 \\
 &=
 \frac{1+\theta_0\theta_1}2,
 \label{eq:phU0(0)}
\end{align}
where the first equality comes from the definition of $U'_0$
and the fact that $U_0$ and $U_1$ are mutually independent.
The above yields
\begin{align}
 P_{U'_0}(1)
 &=
 1-P_{U'_0}(0)
 \notag
 \\
 &=
 \frac{1-\theta_0\theta_1}2.
 \label{eq:phU0(1)}
\end{align}
From (\ref{eq:phU0(0)}) and (\ref{eq:phU0(1)}), we have
\begin{equation}
 P_{U'_0}(u'_0)
 =
 \frac{1\mp_{u'_0}\theta_0\theta_1}2,
\end{equation}
where $\mp_{u'_0}$ is the bipolar-binary conversion of $u'_0$.
Let $\theta'_0$ be defined as
\begin{equation}
 \theta'_0
 \equiv
 P_{U'_0}(0) - P_{U'_0}(1).
 \label{eq:theta'0-def}
\end{equation}
From (\ref{eq:phU0(0)})--(\ref{eq:theta'0-def}), we have
\begin{equation}
 \theta'_0
 =
 \theta_1\theta_0.
 \label{eq:theta'0}
\end{equation}
It should be noted that, since symmetric parametrization is a binary
version of the Fourier transform of
the probability distribution~\cite[Definitions 24 and 25]{MT14},
the right hand side of (\ref{eq:theta'0}) corresponds to 
the Fourier transform of the convolution.

We have
\begin{align}
 P_{U'_0U'_1}(u'_0,0)
 &=
 P_{U_0}(u'_0)P_{U_1}(0)
 \notag
 \\
 &=
 \frac{1\mp_{u'_0}\theta_0}2 \cdot \frac{1+\theta_1}2
 \notag
 \\
 &=
 \frac{1\mp_{u'_0}\theta_0\theta_1+\theta_1\mp_{u'_0}\theta_0}4,
\end{align}
where the first equality comes from the definition of $U'_0$ and $U'_1$,
and the fact that $U_0$ and $U_1$ are mutually independent.
Then we have
\begin{align}
 P_{U'_1|U'_0}(0|u'_0)
 &=
 \frac{P_{U'_1U'_0}(0,u'_0)}
 {P_{U'_0}(u'_0)}
 \notag
 \\*
 &=
 \frac{1+[\theta_1\mp_{u'_0}\theta_0]/[1\mp_{u'_0}\theta_1\theta_0]}
 2
 \label{eq:phU1=0}
 \\
 P_{U'_1|U'_0}(1|u'_0)
 &=
 1-P_{U'_1|U'_0}(0|u'_0)
 \notag
 \\
 &=
 \frac{1-[\theta_1\mp_{u'_0}\theta_0]/[1\mp_{u'_0}\theta_1\theta_0]}
 2.
 \label{eq:phU1=1}
\end{align}
Let $\theta'_1$ be defined as
\begin{equation}
 \theta'_1
 \equiv
 P_{U'_1|U'_0}(0|u_0) - P_{U'_1|U'_0}(1|u_0).
 \label{eq:theta'1-def}
\end{equation}
From (\ref{eq:phU1=0})--(\ref{eq:theta'1-def}),
we have
\begin{align}
 \theta'_1
 &=
 \frac{\theta_1\mp_{u'_0}\theta_0}{1\mp_{u'_0}\theta_1\theta_0}
 \notag
 \\
 &=
 \frac{\theta_1\mp_{u'_0}\theta_0}{1\mp_{u'_0}\theta'_0},
 \label{eq:htheta1}
\end{align}
where the second equality comes from (\ref{eq:theta'0}).

\section{Successive-Cancellation Decoding}

This section introduces the algorithm of successive-cancellation decoding
based on that introduced in~\cite{TV15}.

We assume that Algorithms \ref{alg:SC}--\ref{alg:updateU}
have access to the number of transforms, $n$,
the frozen bits $u^{(n)}_{\I_1}$,
and the memory space
\begin{align*}
 \ttT
 &\equiv
 \lrb{
  \ttT[k][c^{n-k}]:
  \begin{aligned}
   k
   &\in\{0,\ldots,n\}
   \\
   c^{n-k}
   &\in[0^{n-k}:1^{n-k}]
  \end{aligned}
 }
 \\
 \ttU
 &\equiv
 \lrb{
  \ttU[k][c^{n-k}][b]:
  \begin{aligned}
   k
   &\in\{0,\ldots,n\}
   \\
   c^{n-k}
   &\in[0^{n-k}:1^{n-k}]
   \\
   b
   &\in\{0,1\}
  \end{aligned}
 },
\end{align*}
where $\ttT[k][c^{n-k}]$ is a real number variable,
$\ttU[k][c^{n-k}][b]$ is a binary variable,
and $c^0$ denotes the null string.
It should be noted that
$\ttT$ has $\sum_{k=0}^n2^{n-k}=2^{n+1}-1=2N-1$ variables
and $\ttU$ has $2\sum_{k=0}^n2^{n-k} = 2^{n+2}-2=4N-2$ variables.

In the following, we assume that $\{U^{(0)}_{c^n}\}_{c^n\in[0^n:1^n]}$
is a memoryless source, that is, $P_{U^{(0)}_{[0^n:1^n]}}$ is defined as
\begin{equation*}
 P_{U^{(0)}_{[0^n:1^n]}}\lrsb{u^{(0)}_{[0^n:1^n]}}
 \equiv
 \prod_{c^n\in[0^n:1^n]}P_{U^{(0)}_{c^n}}\lrsb{u^{(0)}_{c^n}},
\end{equation*}
where $\{P_{U^{(0)}_{c^n}}\}_{c^n\in[0^n:1^n]}$ is given depending
on the context.
It should be noted that $\{U^{(0)}_{c^n}\}_{c^n\in[0^n:1^n]}$
is allowed to be non-stationary.
We recursively define $U^{(n)}_{b^n}$ as
\begin{align}
 U^{(k)}_{c^{n-k}b^{k-1}0}
 &\equiv
 U^{(k-1)}_{c^{n-k}0b^{k-1}}\oplus U^{(k-1)}_{c^{n-k}1b^{k-1}}
 \label{eq:minus-k}
 \\
 U^{(k)}_{c^{n-k}b^{k-1}1}
 &\equiv
 U^{(k-1)}_{c^{n-k}1b^{k-1}}
 \label{eq:plus-k}
\end{align}
for given $b^n\in\{0,1\}^n$ and $c^{n-k}\in\{0,1\}^{n-k}$.
This yields $U^{(n)}_{[0^n:1^n]}=U^{(0)}_{[0^n:1^n]}G$,
which is the polar transform of $U^{(0)}_{[0^n:1^n]}$.
The goal of $\updateT(\ttT,\ttU,n,\ttb^n)$ at Line 3 of Algorithm \ref{alg:SC}
is to compute
\begin{equation*}
 \theta^{(n)}_{b^n}
 \equiv
 P_{U^{(n)}_{b^n}|U^{(n)}_{[0^n:b^n)}}
 \lrsb{0\left|u^{(n)}_{[0^n:b^n)}\right.}
 -
 P_{U^{(n)}_{b^n}|U^{(n)}_{[0^n:b^n)}}
 \lrsb{1\left|u^{(n)}_{[0^n:b^n)}\right.}
\end{equation*}
recursively starting from
\begin{equation*}
 \theta^{(0)}_{b^n}
 \equiv
 P_{U^{(0)}_{b^n}}(0)-P_{U^{(0)}_{b^n}}(1).
\end{equation*}
In Algorithm \ref{alg:updateT}, we compute a parameter defined as
\begin{align}
 \theta^{(k)}_{c^{n-k}b^k}
 &\equiv
 P_{U^{(k)}_{c^{n-k}b^k}|U^{(k)}_{c^{n-k}[0^k:b^k)}}
 \lrsb{0\left|u^{(k)}_{c^{n-k}[0^k:b^k)}\right.}
 -P_{U^{(k)}_{c^{n-k}b^k}|U^{(k)}_{c^{n-k}[0^k:b^k)}}
 \lrsb{1\left|u^{(k)}_{c^{n-k}[0^k:b^k)}\right.}
 \label{eq:theta-k}
\end{align}
for each $c^{n-k}$ for a given $b^n\in\{0,1\}^n$.
By using (\ref{eq:theta'0}), (\ref{eq:htheta1}), and (\ref{eq:theta-k}),
we have the relations
\begin{align}
 \theta^{(k)}_{c^{n-k}b^{k-1}0}
 &=
 \theta^{(k-1)}_{c^{n-k}1b^{k-1}}\theta^{(k-1)}_{c^{n-k}0b^{k-1}}
 \label{eq:theta-k0}
 \\
 \theta^{(k)}_{c^{n-k}b^{k-1}1}
 &=
 \frac{\theta^{(k-1)}_{c^{n-k}1b^{k-1}}\mp_u\theta^{(k-1)}_{c^{n-k}0b^{k-1}}}
 {1\mp_u\theta^{(k)_{c^{n-k}b^k0}}}
 \label{eq:theta-k1}
\end{align}
where $\mp_u$ is the bipolar-binary conversion
of $u\equiv u^{(k)}_{c^{n-k}b^{k-1}0}$.
The goal of $\updateU(\ttU,n,\ttb^{n-1})$
at Line 9 of Algorithm \ref{alg:SC}
is to compute $u^{(k)}_{c^{n-k}b^{k-1}0}$ from $u^{(n)}_{[0^n:b^{n-1}0]}$
by using the relations
\begin{align}
 u^{(k-1)}_{c^{n-k}0b^{k-1}}
 &\equiv
 u^{(k)}_{c^{n-k}b^{k-1}0}\oplus u^{(k)}_{c^{n-k}b^{k-1}1}
 \label{eq:u-k0}
 \\
 u^{(k-1)}_{c^{n-k}1b^{k-1}}
 &\equiv
 u^{(k)}_{c^{n-k}b^{k-1}1}
 \label{eq:u-k1}
\end{align}
that come from (\ref{eq:minus-k}) and (\ref{eq:plus-k}),
where we assume that $u^{(n)}_{[0^n:b^{n-1}0]}$ is successfully decoded.
It should be noted that
(\ref{eq:theta-k0}) and (\ref{eq:theta-k1})
correspond to Lines 5 and 7 of Algorithm \ref{alg:updateT}, respectively,
and (\ref{eq:u-k0}) and (\ref{eq:u-k1})
correspond to Lines 2 and 3 of Algorithm \ref{alg:updateU}, respectively,
where we have relations
\begin{align}
 \ttT[k][c^{n-k}]
 &= \theta^{(k)}_{c^{n-k}b^k}
 \label{eq:ttT}
 \\
 \ttU[k][c^{n-k}][b_{k-1}]
 &= u^{(k)}_{c^{n-k}b^k}
 \label{eq:ttU}
\end{align}
after completing Lines 3 and 9 of Algorithm \ref{alg:SC},
respectively.
We show (\ref{eq:ttT}) and (\ref{eq:ttU}) in Section \ref{sec:proof-ttT-ttU}.
Furthermore, Line 7 of Algorithm \ref{alg:SC}
corresponds to the maximum a posteriori probability decision defined as
\begin{equation*}
 \hu_{b^n}
 \equiv \arg\max_{u\in\{0,1\}}
 P_{U^{(n)}_{b^n}|U^{(n)}_{[0^n:b^n)}}(u|\hu_{[0^n:b^n)}).
\end{equation*}

When Algorithm \ref{alg:SC} is used for the decoder of polar source code
which has access to codeword $u^{(n)}_{\I_1}$
and side information vector $y_{[0^n:1^n]}$,
we define
\begin{equation}
 P_{U^{(0)}_{c^n}}(x)\equiv P_{X_{c^n}|Y_{c^n}}(x|y_{c^n})
 \label{eq:sourceP}
\end{equation}
for $x\in\{0,1\}$
and obtain the reproduction $\{\hx_{c^n}\}_{c^n\in[0^n:1^n]}$
defined as
\begin{equation}
 \hx_{c^n}\equiv \ttU[0][c^n][\hb_{-1}],
 \label{eq:hX}
\end{equation}
where $\hb_{-1}$ denotes the null string,

When Algorithm \ref{alg:SC} is to decode a systematic polar channel code,
which has access to channel output vector $y_{[0^n:1^n]}$
and shared vector $u^{(n)}_{\I_1}$,
we define
\begin{equation*}
 P_{U^{(0)}_{c^n}}(x)
 \equiv
 \frac{P_{Y_{c^n}|X_{c^n}}(y_{c^n}|x)P_{X_{c^n}}(x)}
 {\sum_{x'\in\{0,1\}}P_{Y_{c^n}|X_{c^n}}(y_{c^n}|x')P_{X_{c^n}}(x')}
\end{equation*}
for given channel distribution $\{P_{Y_{c^n}|X_{c^n}}\}_{c^n\in[0^n:1^n]}$,
input distribution $\{P_{X_{c^n}}\}_{c^n\in[0^n:1^n]}$,
$x\in\{0,1\}$, and $y_{c^n}\in\Y$,
This yields a reproduction $\{\hx_{c^n}\}_{c^n\in\I'_0}$
defined by (\ref{eq:hX}), where $\I'_0$ is defined by (\ref{eq:I'0}).

When Algorithm \ref{alg:SC} is used in the decoder
of a non-systematic polar channel code,
we have to prepare binary variables $\{\ttM[b^n]\}_{b^n\in\I_0}$
and insert
\begin{equation*}
 \ttM[\ttb^n]\leftarrow \ttU[n][\ttc^0][\ttb_{n-1}]
\end{equation*}
just after the renewal of $\ttU[n][\ttc^0][\ttb_{n-1}]$
(Line 7 of Algorithm \ref{alg:SC}).
This yields reproduction $\hu_{\I_0}$ defined as
\begin{equation*}
 \hu_{b^n}\equiv \ttM[b^n].
\end{equation*}

\begin{algorithm}[t]
 \caption{Successive-cancellation decoder}
 \hspace*{\algorithmicindent}\textbf{Input:}
 $\I_1$, $u^{(n)}_{\I_1}$, $\lrb{P_{U^{(0)}_{b^n}}}_{b^n\in[0^n:1^n]}$
 \label{alg:SC}
 \begin{algorithmic}[1]
  \FFor{$\ttb^n\in[0^n:1^n]$}
  $\ttT[0][\ttb^n]\leftarrow P_{U^{(0)}_{\ttb^n}}(0)-P_{U^{(0)}_{\ttb^n}}(1)$
  \For{$\ttb^n\in[0^n:1^n]$}
  \State $\updateT(\ttT,\ttU,n,\ttb^n)$
  \If{$\ttb^n\in\I_1$}
  \State
  $\ttU[n][\ttc^0][\ttb_{n-1}]\leftarrow u^{(n)}_{\ttb^n}$
  \Else
  \State
  $\ttU[n][\ttc^0][\ttb_{n-1}]\leftarrow
  \begin{cases}
   0
   &\text{if}\ \ttT[n][0]>0
   \\
   1
   &\text{if}\ \ttT[n][0]<0
   \\
   \text{$0$ or $1$}
   &\text{if}\ \ttT[n][0]=0
  \end{cases}$
  \EndIf
  \IIf{$\ttb_{n-1}=1$}
  $\updateU(\ttU,n,\ttb^{n-1})$
  \EndFor
 \end{algorithmic}
\end{algorithm}

\begin{algorithm}[t]
 \caption{$\updateT(\ttT,\ttU,\ttk,\ttb^{\ttk})$}
 \label{alg:updateT}
 \begin{algorithmic}[1]
  \IIf{$k=0$}
  \Return
  \IIf{$\ttb_{\ttk-1}=0$}
  $\updateT(\ttT,\ttU,\ttk-1,\ttb^{\ttk-1})$
  \For{$\ttc^{n-k}\in[0^{n-\ttk}:1^{n-\ttk}]$}
  \If{$\ttb_{\ttk-1}=0$}
  \State
  $\ttT[\ttk][\ttc^{n-\ttk}]
  \leftarrow 
  \ttT[\ttk-1][\ttc^{n-\ttk}1]
  \cdot
  \ttT[\ttk-1][\ttc^{n-\ttk}0]$
  \Else
  \State
  $\ttT[\ttk][\ttc^{n-\ttk}]
  \leftarrow 
  \frac{\ttT[\ttk-1][\ttc^{n-\ttk}1]\mp_{\ttu}\ttT[\ttk-1][\ttc^{n-\ttk}0]}
  {1\mp_{\ttu}\ttT[\ttk][\ttc^{n-\ttk}]}$,
  where $\ttu\equiv\ttU[\ttk][\ttc^{n-\ttk}][0]$
  \EndIf
  \EndFor
 \end{algorithmic}
\end{algorithm}

\begin{algorithm}[t]
 \caption{$\updateU(\ttU,\ttk,\ttb^{\ttk-1})$}
 \label{alg:updateU}
 \begin{algorithmic}[1]
  \For{$\ttc^{n-\ttk}\in[0^{n-\ttk}:1^{n-\ttk}]$}
  \State
  \!\!$\ttU[\ttk-1][\ttc^{n-\ttk}0][\ttb_{\ttk-2}]
  \leftarrow
  \ttU[\ttk][\ttc^{n-\ttk}][0]\oplus\ttU[\ttk][\ttc^{n-\ttk}][1]$
  \State
  \!\!$\ttU[\ttk-1][\ttc^{n-\ttk}1][\ttb_{\ttk-2}]
  \leftarrow
  \ttU[\ttk][\ttc^{n-\ttk}][1]$
  \EndFor
  \IIf{$\ttb_{\ttk-2}=1$}
  $\updateU(\ttU,\ttk-1,\ttb^{\ttk-2})$
 \end{algorithmic}
\end{algorithm}

\begin{rem}
\label{rem:systematic}
When Algorithm \ref{alg:SC} is applied to a binary erasure channel,
we can assume that $\ttT[\ttk][\ttc^{n-\ttk}]$
takes a value in $\{-1,0,1\}$,
where
$\ttT[0][\ttb^n]\leftarrow P_{U^{(0)}_{\ttb^n}}(0)-P_{U^{(0)}_{\ttb^n}}(1)$
in Line 1 of Algorithm \ref{alg:SC} can be replaced by
\begin{equation*}
 \ttT[0][\ttb^n]\leftarrow
 \begin{cases}
  1
  &\text{if $y_{\ttb^n}=0$}
  \\
  0
  &\text{if $y_{\ttb^n}$ is the erasure symbol}
  \\
  -1
  &\text{if $y_{\ttb^n}=1$}.
 \end{cases}
\end{equation*}
for given channel output $y_{[0^n:1^n]}$.
We can improve Algorithm~\ref{alg:updateT} as described in
Appendix~\ref{sec:updateT}.
\end{rem}

\section{Successive-Cancellation List Decoding}

This section introduces an algorithm for
the successive-cancellation list decoding.
It is based on that introduced in \cite{TV15}.
It should be noted that
we use a fixed-addressing memory space
instead of the stacking memory space approach used in \cite{TV15}.
Since the size of memory space for the computation of conditional probability
is around half that used in \cite{TV15},
the time complexity of our algorithm is around half
that mentioned in \cite{TV15}.

We assume that Algorithms \ref{alg:updateT}--\ref{alg:magnifyP}
have access to
the number of transforms, $n$, the list size $L$,
the frozen bits $u^{(n)}_{\I_1}$, and
the memory space
$\{\ttT[\lambda]\}_{\lambda=0}^{L-1}$,
$\{\ttU[\lambda]\}_{\lambda=0}^{L-1}$,
$\{\ttP[\lambda]\}_{\lambda=0}^{2L-1}$,
$\{\Active[\lambda]\}_{\lambda=0}^{2L-1}$,
where $\ttT[\lambda]$ and $\ttU[\lambda]$
are accessed by Algorithms \ref{alg:updateT} and \ref{alg:updateU},
$\{\ttP[\lambda]\}_{\lambda=0}^{2L-1}$ are real number variables,
and $\{\Active[\lambda]\}_{\lambda=0}^{2L-1}$ are binary variables.
After Algorithm \ref{alg:SCL} concludes,
the results are stored in $\{\ttU[\lambda]\}_{\lambda=0}^{L-1}$
and $\{\ttP[\lambda]\}_{\lambda=0}^{2L-1}$ satisfying
\begin{align}
 \frac{\ttP[\lambda]}{2^N}
 &= 
 \prod_{b^n\in[0^n:1^n]}
 P_{U^{(n)}_{b^n}|U^{(n)}_{[0^n:b^n)}}
 \lrsb{\hu^{(n)}_{b^n}(\lambda)\left|\hu^{(n)}_{[0^n:b^n)}(\lambda)\right.}
 \notag
 \\
 &=
 P_{U^{(n)}_{[0^n:1^n]}}\lrsb{\hu^{(n)}_{[0^n:1^n]}(\lambda)}
 \label{eq:Plambda-1^n}
\end{align}
and
\begin{align*}
 \ttU[\lambda][0][c^n][b_{-1}]
 &=
 \hu^{(0)}_{c^n}(\lambda)
 \\
 \hu^{(n)}_{[0^n:1^n]}(\lambda)
 &=
 \hu^{(0)}_{[0^n:1^n]}(\lambda)G,
\end{align*}
where $\hu^{(n)}_{[0^n:1^n]}(\lambda)$ is the $\lambda$-th surviving path.
It should be noted that, at Line 5 of Algorithm \ref{alg:prunePath},
we select paths $\hu^{(n)}_{[0^n:b^n]}$
that have the $L$ largest probability
\begin{align}
 \frac{\ttP[\lambda]}{2^{|[0^n:b^n]|}}
 &= 
 \prod_{d^n\in[0^n:b^n]}
 P_{U^{(n)}_{d^n}|U^{(n)}_{[0^n:d^n)}}
 \lrsb{\hu^{(n)}_{d^n}(\lambda)\left|\hu^{(n)}_{[0^n:d^n)}(\lambda)\right.}
 \notag
 \\*
 &=
 P_{U^{(n)}_{[0^n:b^n]}}\lrsb{\hu^{(n)}_{[0^n:b^n]}(\lambda)},
 \label{eq:Plambda-b^n}
\end{align}
where $\hu^{(n)}_{[0^n:b^n]}(\lambda)$ is the $\lambda$-th surviving path.
We show (\ref{eq:Plambda-1^n}) and (\ref{eq:Plambda-b^n}) in Section
\ref{sec:proof-Plambda}.

When Algorithm \ref{alg:SCL} is used in the decoder of polar source code
that has access to
the codeword $u_{\I_1}$
and side information vector
$y_{[0^n:1^n]}$,
we define $P_{U^{(0)}_{b^n}}$ by (\ref{eq:sourceP})
and obtain reproduction $\{\hx_{c^n}(l)\}_{c^n\in[0^n:1^n]}$ defined as
\begin{equation}
 \hx_{c^n}(l)\equiv \ttU[l][0][c^n][b_{-1}],
 \label{eq:scl-source-reproduction}
\end{equation}
where
\begin{equation}
 l\equiv\arg\max_{\lambda} \ttP[\lambda].
 \label{eq:lmax}
\end{equation}
When the outer parity check function $\parity$ generates
an extension $\bs\equiv\parity(x^n)$ to the codeword $u_{\I_1}$,
the corresponding reproduction is defined as (\ref{eq:scl-source-reproduction})
for an $l$ satisfying $\parity\lrsb{\{\hx_{c^n}(l)\}_{c^n\in[0^n:1^n]}}=\bs$.

When Algorithm \ref{alg:SCL} is used in the decoder of
systematic polar channel code
that has access to channel output vector $y_{[0^n:1^n]}$,
and shared vector $u_{\I_1}$,
we obtain reproduction $\{\hx_{c^n}(l)\}_{c^n\in\I'_0}$ defined by
(\ref{eq:scl-source-reproduction}) and (\ref{eq:lmax}), where $\I'_0$ is defined by (\ref{eq:I'0}).
When we use the outer parity check function $\parity$ (e.g. polar code with CRC \cite{TV15})
with check vector $\bs$
satisfying $\bs=\parity(x^n)$ for all channel inputs $x^n$,
the resulting reproduction $\{\hx_{c^n}(l)\}_{c^n\in\I'_0}$ is defined as
(\ref{eq:scl-source-reproduction})
for an $l$ satisfying $\parity\lrsb{\{\hx_{c^n}(l)\}_{c^n\in[0^n:1^n]}}=\bs$.

When Algorithm \ref{alg:SCL} is used in the decoder of
non-systematic polar channel code,
we have to prepare binary variables
$\{\ttM[\lambda][b^n]\}_{\lambda\in\{0,\ldots L-1\}, b^n\in\I_0}$
and insert
\begin{align*}
 &\ttM[\ttl][\ttb^n]\leftarrow \ttU[\ttl][n][\ttc^0][\ttb_{n-1}]
 \\
 &\ttM[\ttL+\ttl][b^n]\leftarrow \ttU[\ttL+\ttl][n][\ttc^0][\ttb_{n-1}]
\end{align*}
just after the renewal of $\ttU[\ttl][n][\ttc^0][\ttb_{n-1}]$
and $\ttU[\ttL+\ttl][n][\ttc^0][\ttb_{n-1}]$, respectively
(Lines 5 and 6 of Algorithm \ref{alg:splitPath},
 and Lines 8, 11, 15, and 26 of Algorithm \ref{alg:prunePath}).
This yields reproduction
$\{\hu_{b^n}(l)\}_{b^n\in\I_0}$ defined as
\begin{equation}
 \hu_{b^n}(l)\equiv \ttM[l][b^n],
 \label{eq:scl-channel-reproduction-NS}
\end{equation}
where $l$ is defined by (\ref{eq:lmax}).
When we use the outer parity check function $\parity$ (e.g. polar code with CRC \cite{TV15})
with check vector $\bs$
satisfying $\bs=\parity(x^n)$ for all channel input $x^n$,
the resulting reproduction $\{\hu_{b^n}(l)\}_{b^n\in\I_0}$
of the non-systematic code is defined as
(\ref{eq:scl-channel-reproduction-NS})
for an $l$ such that 
corresponding channel input $\{\hx_{c^n}(l)\}_{c^n\in[0^n:1^n]}$
defined by (\ref{eq:scl-source-reproduction}) 
satisfies $\parity\lrsb{\{\hx_{c^n}(l)\}_{c^n\in[0^n:1^n]}}=\bs$.

\begin{algorithm}[t]
 \caption{Successive-cancellation list decoder}
 \hspace*{\algorithmicindent}\textbf{Input:}
 $\I_1$, $u^{(n)}_{\I_1}$, $\lrb{P_{U^{(0)}_{b^n}}}_{b^n\in[0^n:1^n]}$,
 $L$
 \label{alg:SCL}
 \begin{algorithmic}[1]
  \State $\ttL\leftarrow1$
  \FFor{$\ttb^n\in[0^n:1^n]$}
  $\ttT[0][0][\ttb^n]
  \leftarrow P_{U^{(0)}_{\ttb^n}}(0)-P_{U^{(0)}_{\ttb^n}}(1)$
  \State $\ttP[0]\leftarrow 1$
  \For{$\ttb^n\in[0^n:1^n]$}
  \FFor{$\ttl\in\{0,...,\ttL-1\}$}
  $\updateT(\ttT[\ttl],\ttU[\ttl],n,\ttb^n)$
  \If{$\ttb^n\in\I_1$}
  \State $\extendPath(\ttb^n,\ttL)$
  \Else
  \If{$2\cdot\ttL\leq L$}
  \State $\splitPath(\ttb^n,\ttL)$
  \State $\ttL\leftarrow 2\cdot\ttL$
  \Else
  \State $\prunePath(\ttb^n,\ttL)$
  \State $\ttL\leftarrow L$
  \EndIf
  \EndIf
  \State $\magnifyP(\ttL)$
  \If{$\ttb_{n-1}=1$}
  \FFor{$\ttl\in\{0,...,\ttL-1\}$}
  $\updateU(\ttU[\ttl],n,\ttb^{n-1})$
  \EndIf
  \EndFor
 \end{algorithmic}
\end{algorithm}

\begin{algorithm}[t]
 \caption{$\extendPath(\ttb^n,\ttL)$}
 \label{alg:extendPath}
 \begin{algorithmic}[1]
  \For{$\ttl\in\{0,...,\ttL-1\}$}
  \State
  $\ttP[\ttl]\leftarrow \ttP[\ttl]\cdot(1\mp_{\ttu}\ttT[\ttl][n][\ttb^n])$,
  where $\ttu\equiv u^{(n)}_{\ttb^n}$
  \State $\ttU[\ttl][n][c^0][\ttb_{n-1}]\leftarrow u^{(n)}_{\ttb^n}$
  \EndFor
 \end{algorithmic}
\end{algorithm}

\begin{algorithm}[t]
 \caption{$\splitPath(\ttb^n,\ttL)$}
 \label{alg:splitPath}
 \begin{algorithmic}[1]
  \For{$\ttl\in\{0,\ldots,\ttL-1\}$}
  \State $\ttP[\ttL+\ttl]\leftarrow \ttP[\ttl]\cdot(1-\ttT[\ttl][n][\ttb^n])$
  \State $\ttP[\ttl]\leftarrow \ttP[\ttl]\cdot(1+\ttT[\ttl][n][b^n])$
  \State $\copyPath(\ttL+\ttl,\ttl)$
  \State $\ttU[\ttl][n][\ttc^0][\ttb_{n-1}]\leftarrow 0$
  \State $\ttU[\ttL+\ttl][n][\ttc^0][\ttb_{n-1}]\leftarrow 1$
  \EndFor
 \end{algorithmic}
\end{algorithm}

For completeness,
we introduce an algorithm for Line 5 of Algorithm \ref{alg:prunePath} 
in Appendix \ref{sec:pruePath}.

\begin{algorithm}[t]
 \caption{$\prunePath(\ttb^n,\ttL)$}
 \label{alg:prunePath}
 \begin{algorithmic}[1]
  \For{$\ttl\in\{0,\ldots,\ttL-1\}$}
  \State $\ttP[\ttL+\ttl]\leftarrow \ttP[\ttl]\cdot(1-\ttT[\ttl][n][\ttb^n])$
  \State $\ttP[\ttl]\leftarrow \ttP[\ttl]\cdot(1+\ttT[\ttl][n][\ttb^n])$
  \EndFor
  \State Define $\{\Active[\ttl]\}_{\ttl=0}^{2\cdot\ttL-1}$
  such that $\Active[\ttl]=1$ iff $\ttP[\ttl]$ is one of the $L$ largest values
  of $\{\ttP[\ttl]\}_{\ttl=0}^{2\cdot\ttL-1}$ (ties are broken arbitrarily).
  \For{$\ttl\in\{0,\ldots,\ttL-1\}$}
  \If{$\Active[\ttl]=1$}
  \State $\ttU[\ttl][n][\ttc^0][\ttb_{n-1}]\leftarrow 0$
  \If{$\Active[\ttL+\ttl]=1$ and $\ttL+\ttl<L$}
  \State $\copyPath(\ttL+\ttl,\ttl)$
  \State $\ttU[\ttL+\ttl][n][\ttc^0][\ttb_{n-1}]\leftarrow 1$
  \EndIf
  \ElsIf{$\Active[\ttL+\ttl]=1$}
  \State $\ttP[\ttl]\leftarrow\ttP[\ttL+\ttl]$
  \State $\ttU[\ttl][n][\ttc^0][\ttb_{n-1}]\leftarrow 1$
  \State $\Active[\ttl]\leftarrow1$
  \State $\Active[\ttL+\ttl]\leftarrow0$
  \EndIf
  \EndFor
  \State $\ttl'\leftarrow0$
  \For{$\ttl\in\{L,\ldots,2\cdot\ttL-1\}$}
  \If{$\Active[\ttl]=1$}
  \WWhile{$\Active[\ttl]=1$}
  $\ttl'\leftarrow \ttl'+1$
  \State $\ttP[\ttl']\leftarrow\ttP[\ttl]$
  \State $\copyPath(\ttl',\ttl-\ttL)$
  \State $\ttU[\ttl][n][\ttc^0][\ttb_{n-1}]\leftarrow 1$
  \State $\ttl'\leftarrow \ttl'+1$
  \EndIf
  \EndFor
 \end{algorithmic}
\end{algorithm}

\begin{algorithm}[t]
 \caption{$\copyPath(\ttl',\ttl)$}
 \label{alg:copyPath}
 \begin{algorithmic}[1]
  \For{$\ttk\in\{0,\ldots,n\}$}
  \For{$\ttc^{\ttk}\in\{0^{\ttk},\ldots,1^{\ttk}\}$}
  \State $\ttT[\ttl'][\ttk][\ttc^{\ttk}] \leftarrow\ttT[\ttl][\ttk][c^{\ttk}]$
  \For{$\ttb\in\{0,1\}$}
  \State $\ttU[\ttl'][\ttk][\ttc^{\ttk}][\ttb]
  \leftarrow\ttU[\ttl][\ttk][\ttc^{\ttk}][\ttb]$
  \EndFor
  \EndFor
  \EndFor
 \end{algorithmic}
\end{algorithm}

\begin{algorithm}[t]
 \caption{$\magnifyP(\ttL)$}
 \label{alg:magnifyP}
 \begin{algorithmic}[1]
  \State $\maxP\leftarrow 0$
  \For{$\ttl\in\{0,\ldots,\ttL-1\}$}
  \IIf{$\maxP<\ttP[\ttl]$}
  $\maxP\leftarrow\ttP[\ttl]$
  \EndFor
  \FFor{$\ttl\in\{0,\ldots,\ttL-1\}$}
  $\ttP[\ttl]\leftarrow\ttP[\ttl]/\maxP$
 \end{algorithmic}
\end{algorithm}

\begin{rem}
Line 17 of Algorithm \ref{alg:SCL}
is unnecessary if we use the infinite precision real number variables.
We assumed the use of the finite precision (floating point)
real number variables
to prevent $\ttP[\lambda]$ from vanishing as $b^n$ increases.
We can skip Line 17 of Algorithm \ref{alg:SCL}
and Line 2 of Algorithm \ref{alg:extendPath}
while $b^n\in\I_1$ is satisfied continuously from the beginning ($b^n=0^n$).
It should be noted that this type of technique
is used in \cite[Algorithm 10, Lines 20--25]{TV15},
where this technique is repeated $Nn$ times.
In contrast, Algorithm \ref{alg:SCL} uses this technique
outside the renewal of parameters $\{\ttT[\lambda]\}_{\lambda=0}^{L-1}$
(Algorithm \ref{alg:updateT}),
where $\magnifyP(\Lambda)$ is repeated $N$ times.
\end{rem}

\begin{rem}
When we assume that $L$ is a power of $2$,
$\ttL=L$ is always satisfied at Line 13 of Algorithm \ref{alg:SCL}.
Accordingly, we can omit Line 14 of Algorithm \ref{alg:SCL}
and Lines 9--12 of Algorithm \ref{alg:prunePath}
because $\ttL+l\geq L$ is always satisfied.
\end{rem}

\section{Proofs}

\subsection{Proof of (\ref{eq:ttT}) and (\ref{eq:ttU})}
\label{sec:proof-ttT-ttU}
Here, we check that we can compute $u^{(k)}_{b^k0}$ from $u^{(n)}_{[0^n:b^n)}$.
We introduce the following theorems.
In the proof of theorems, we write $d^n<b^n$
when the corresponding integers satisfy the same relation.

\begin{thm}
\label{thm:updateU}
For a given $b^n\equiv(b_0,b_1,\ldots,b_{n-1})$, we have
\begin{equation}
 \ttU[k][c^{n-k}][b_{k-1}] = u^{(k)}_{c^{n-k}b^k}
 \label{eq:ttUk}
\end{equation}
for all $k$ and $c^{n-k}\in\{0,1\}^{n-k}$ after the operations
\begin{gather}
 \ttU[n][c^0][0]\leftarrow u^{(n)}_{d^{n-1}0}
 \label{eq:ttUn0}
 \\
 \ttU[n][c^0][1]\leftarrow u^{(n)}_{d^{n-1}1}
 \label{eq:ttUn1}
 \\
 \updateU(\ttU,n,d^{n-1})
 \label{eq:updateU}
\end{gather}
employed for each $d^{n-1}\in[0^n:b^{n-1}]$.
In particular, after the operations (\ref{eq:ttUn0})--(\ref{eq:updateU})
for each $d^{n-1}\in[0^n:1^{n-1}]$,
\begin{equation}
 \ttU[0][c^n][d_{-1}]= u^{(0)}_{c^n}
 \label{eq:ttU0}
\end{equation}
for all $c^n\in[0^n:1^n]$.
\end{thm}
\begin{IEEEproof}
 For a given $b^n\in[0^n:1^n]$, we have
 \begin{equation*}
  \ttU[n][c^0][b_{n-1}] = u^{(n)}_{b^n} = u^{(n)}_{c^0b^n}
 \end{equation*}
 after the operations (\ref{eq:ttUn0}) and (\ref{eq:ttUn1}).
 
 From Line 4 of Algorithm \ref{alg:updateU},
 $\updateU(\ttU,k,b^{k-1})$ is called only when
 $(b_{k-1}, \ldots, b_{n-1}) = 1^{n-k+1}$.
 Let us assume that $(b_k,\ldots,b_{n-1})=1^{n-k}$.
 Since $b^k01^{n-k-1}<b^n$,
 we have the fact that $\updateU(k+1,b^k)$ is called
 and $\ttU[k][c^{n-k-1}0][0]$ and $\ttU[k][c^{n-k-1}1][0]$ are defined.
 Here, let us assume that $\ttU[k][c^{n-k}][b]=u^{(k)}_{c^{n-k}b^{k-1}b}$
 for all $c^{n-k}$ and $b\in\{0,1\}$.
 Then we have
 \begin{align}
  \ttU[k-1][c^{n-k}0][b_{k-2}]
  &=
  \ttU[k][c^{n-k}][0]\oplus \ttU[k][c^{n-k}][1]
  \notag
  \\
  &=
  u^{(k)}_{c^{n-k}b^{k-1}0}\oplus u^{(k)}_{c^{n-k}b^{k-1}1}
  \notag
  \\
  &=
  u^{(k-1)}_{c^{n-k}0b^{k-1}}\oplus u^{(k-1)}_{c^{n-k}1b^{k-1}}
  \oplus u^{(k-1)}_{c^{n-k}1b^{k-1}}
  \notag
  \\
  &=
  u^{(k-1)}_{c^{n-k}0b^{k-1}},
 \end{align}
 where the third equality comes from (\ref{eq:u-k0}) and (\ref{eq:u-k1}). 
 In addition, we have
 \begin{align}
  \ttU[k-1][c^{n-k}1][b_{k-2}]
  &=
  \ttU[k][c^{n-k}][1]
  \notag
  \\
  &=
  u^{(k)}_{c^{n-k}b^{k-1}1}
  \notag
  \\
  &=
  u^{(k-1)}_{c^{n-k}1b^{k-1}},
 \end{align}
 where the last equality comes from (\ref{eq:u-k1}). 
 The above yields
 \begin{equation*}
  \ttU[k-1][c^{n-k+1}][b_{k-2}]
  =
  u^{(k-1)}_{c^{n-k+1}b^{k-1}},
 \end{equation*}
 for all $c^{n-k+1}$.
 Induction yields the relation (\ref{eq:ttUk}) for all $k$ and $c^{n-k}$
 for a given $b^n$.
 By letting $k=0$ and $b^n=1^n$, we have
 the fact that $\updateU(\ttU,0,b_{-1})$ is called
 and $\ttU[n][c^n][b_{-1}]$ satisfies (\ref{eq:ttU0}) for all $c^n$.
\end{IEEEproof}

\begin{thm}
\label{eq:updateT}
Assume that
\begin{equation}
 \ttT[0][c^n] = \theta^{(0)}_{c^n}
 \label{eq:ttT0}
\end{equation}
for all $c^n\in[0^n:1^n]$.
Then, for a given $b^n\equiv(b_0,b_1,\ldots,b_{n-1})$,
we have
\begin{equation}
 \ttT[k][c^{n-k}] = \theta^{(k)}_{c^{n-k}b^k}
 \label{eq:ttTk}
\end{equation}
for all $k$ and $c^{n-k}\in\{0,1\}^{n-k}$ after the operations
\begin{gather*}
 \updateT(\ttT,n,d^{n-1})
 \\
 \ttU[n][c^0][0]\leftarrow u^{(n)}_{d^{n-1}0}
 \\
 \ttU[n][c^0][1]\leftarrow u^{(n)}_{d^{n-1}1}
 \\
 \updateU(\ttU,n,d^{n-1})
\end{gather*}
for each $d^{n-1}\in[0^n:b^{n-1})$ and
\begin{equation*}
 \updateT(\ttT,n,b^{n-1}).
\end{equation*}
\end{thm}
\begin{IEEEproof}
 We have the fact that $\updateT(\ttT,k-1,b^{k-1})$ is called
 only when $(b_k,\ldots,b_{n-1})=0^{n-k}$.
 Let $b^0$ denote the null string.
 
 Let us assume that (\ref{eq:ttTk}) is satisfied for all
 $k\in\{1,\ldots,n-1\}$ and $c^{n-k}$ and
 \begin{align}
  \ttT[k-1][c^{n-k+1}] 
  &=
  \theta^{(k-1)}_{c^{n-k+1}b^{k-1}}
  \label{eq:ttT(k-1)}
 \end{align}
 for all $c^{n-1}$ and $b^n$,
 where this equality is satisfied when $k=1$ from assumption (\ref{eq:ttT0}).
 
 Assume that $b_{k-1}=0$.
 Since $b^{k-2}0^{n-k+2}<b^n$,
 then $\updateT(\ttT,k-1,b^{k-1})$ is called and
 \begin{align}
  \ttT[k][c^{n-k}] 
  &= 
  \ttT[k-1][c^{n-k}1]\cdot\ttT[k-1][c^{n-k}0]
  \notag
  \\
  &=
  \theta^{(k-1)}_{c^{n-k}1b^{k-1}}\theta^{(k-1)}_{c^{n-k}0b^{k-1}}
  \notag
  \\
  &=
  \theta^{(k)}_{c^{n-k}b^{k-1}0}
  \label{eq:ttT(k)0}
 \end{align}
 for all $c^{n-1}$,
 the first equality comes from Line 5 of Algorithm~\ref{alg:updateU},
 the second equality comes from (\ref{eq:ttT(k-1)}),
 and the last equality comes from (\ref{eq:theta-k0}).
 
 Assume that $b_{k-1}=1$.
 Since $b^{k-2}0^{n-k+2}<b^n$,
 $\updateT(\ttT,k-1,b^{k-1})$ is called and we have (\ref{eq:ttT(k)0}).
 Furthermore, since $b^{k-2}01^{n-k+1}<b^n$,
 then $\updateU(\ttU,k,b^{k-1})$ is called and we have
 \begin{equation}
  \ttU[k][c^{n-k}][0]=u^{(k)}_{c^{n-k}b^{k-1}0}
  \label{eq:ttU(1)0}
 \end{equation}
 from Theorem \ref{thm:updateU}.
 Then we have
 \begin{align}
  \ttT[k][c^{n-k}] 
  &= 
  \frac{\ttT[k-1][c^{n-k}1]
   \mp_u\ttT[k-1][c^{n-k}0]}
  {1\mp_u \theta^{(k)}_{c^{n-k}b^{k-1}0}}
  \notag
  \\
  &=
  \frac{\theta^{(k-1)}_{c^{n-k}1b^{k-1}}\mp_u\theta^{(k-1)}_{c^{n-k}0b^{k-1}}}
  {1\mp_u \theta^{(k)}_{c^{n-k}b^{k-1}0}}
  \notag
  \\
  &=
  \theta^{(k)}_{c^{n-k}b^{k-1}1}
  \label{eq:ttT(k)1}
 \end{align}
 for all $c^{n-k}$,
 where
 \begin{align}
  u
  &\equiv
  \ttU[k][c^{n-k}][0]
  \notag
  \\
  &=
  u^{(k)}_{c^{n-k}b^{k-1}0}
 \end{align}
 from (\ref{eq:ttU(1)0}),
 where the first equality comes from Line 7 of Algorithm~\ref{alg:updateU}
 and (\ref{eq:ttT(k)0}),
 the second equality comes from (\ref{eq:ttT(k-1)}),
 and the last equality comes from (\ref{eq:theta-k1}).
 
 From (\ref{eq:ttT(k)0}) and (\ref{eq:ttT(k)1}),
 we have (\ref{eq:ttTk}) for all $c^{n-1}$ and $b^n$ by induction.
\end{IEEEproof}

\subsection{Proof of (\ref{eq:Plambda-1^n}) and (\ref{eq:Plambda-b^n})}
\label{sec:proof-Plambda}

Here, we show (\ref{eq:Plambda-b^n}) by proving the folloing theorem,
where (\ref{eq:Plambda-1^n}) is shown by letting $b^n\equiv 1^n$.

\begin{thm}
Let $\hu^{(n)}_{[0^n:b^n]}(\lambda)$ be the $\lambda$-th surviving path
after employing one of Algorithms \ref{alg:extendPath}--\ref{alg:prunePath}
(at Line 17 of Algoirthm~\ref{alg:SCL}).
We have
\begin{align}
 \frac{\ttP[\lambda]}{2^{|[0^n:b^n]|}}
 &= 
 \prod_{d^n\in[0^n:b^n]}
 P_{U^{(n)}_{d^n}|U^{(n)}_{[0^n:d^n)}}
 \lrsb{\hu^{(n)}_{d^n}(\lambda)\left|\hu^{(n)}_{[0^n:d^n)}(\lambda)\right.}
 \notag
 \\*
 &=
 P_{U^{(n)}_{[0^n:b^n]}}\lrsb{\hu^{(n)}_{[0^n:b^n]}(\lambda)}.
\end{align}
\end{thm}
\begin{IEEEproof}
 After employing one of Algorithms \ref{alg:extendPath}--\ref{alg:prunePath},
 we have the substitution
 \begin{equation*}
  P[\lambda] \leftarrow P[\lambda]\cdot(1\mp_{\hu}\ttT[\lambda][n][b^n]),
 \end{equation*}
 where 
 \begin{equation}
  \hu(\lambda)\equiv \hu^{(n)}_{b^n}(\lambda)
  \label{eq:hu(lambda)}
 \end{equation}
 is the $b^n$-th symbol of the $\lambda$-th surviving binary path
 $\hu^{(n)}_{[0^n:b^n]}(\lambda)$.
 Then we have
 \begin{align}
  \frac{P[\lambda]}{2^{|[0^n:b^n]|}}
  &=
  \prod_{d^n\in[0^n:b^n]}
  \frac{
   1\mp_{\hu(\lambda)}\ttT[\lambda][n][c^0]
  }2
  \notag
  \\
  &=
  \prod_{d^n\in[0^n:b^n]}
  \frac{
   1\mp_{\hu(\lambda)}\theta^{(n)}_{d^n}(\lambda)
  }2
  \notag
  \\
  &=
  \prod_{d^n\in[0^n:b^n]}
  P_{U^{(n)}_{d^n}|U^{(n)}_{[0^n:d^n)}}
  \lrsb{\hu^{(n)}_{d^n}(\lambda)\left|\hu^{(n)}_{[0^n:d^n)}(\lambda)\right.}
  \notag
  \\
  &=
  P_{U^{(n)}_{[0^n:b^n]}}
  \lrsb{\hu^{(n)}_{[0^n:b^n]}(\lambda)},
 \end{align}
 where
 \begin{align}
  \theta^{(n)}_{d^n}(\lambda)
  &\equiv
  P_{U^{(n)}_{d^n}|U^{(n)}_{[0^n:d^n)}}
  \lrsb{0\left|\hu^{(n)}_{[0^n:d^n)}(\lambda)\right.}
  -
  P_{U^{(n)}_{d^n}|U^{(n)}_{[0^n:d^n)}}
  \lrsb{1\left|\hu^{(n)}_{[0^n:d^n)}(\lambda)\right.},
  \label{eq:proof-Plambda-theta}
 \end{align}
 the first equality is shown by induction,
 the second equality comes from Theorem \ref{eq:updateT},
 and the third equality comes from (\ref{eq:hu(lambda)})
 and (\ref{eq:proof-Plambda-theta}).
\end{IEEEproof}

\appendix
\subsection{Algorithm for Polar Transform}
\label{sec:polar-transform}
This section introduces the polar transform
defined by (\ref{eq:minus-k}) and (\ref{eq:plus-k}).
We assume that the following algorithm have access to memory space
\begin{equation*}
 \lrb{
  \ttu[k][b^n]:
  \begin{aligned}
   k&\in\{0,1\}
   \\
   b^n&\in[0^n:1^n]
  \end{aligned}
 }
\end{equation*}
and function $\lsb(k)$ outputs the least significant bit of $k$,
which equals $k\mod 2$.
The result is stored in $\{\ttu[\lsb(n)][b^n]\}_{b^n\in[0^n:1^n]}$.
It should be noted that, from the relation $G=G^{-1}$,
we have the fact that $\uu=\xx G$ is equivalent to $\xx=\uu G$.
This implies that we can obtain $\xx$ from $\uu$
so that $\xx=\uu G^{-1}=\uu G$ is satisfied.

\begin{algorithm}[H]
 \caption{Polar transform}
 \hspace*{\algorithmicindent}\textbf{Input:} $u^{(0)}_{[0^n:1^n]}$
 \begin{algorithmic}[1]
  \FFor{$\ttb^n\in[0^n:1^n]$}
  $\ttu[0][\ttb^n]\leftarrow u^{(0)}_{b^n}$
  \For{$\ttk\in\{0,\ldots,n-1\}$}
  \For{$\ttc^{n-\ttk-1}\in[0^{n-\ttk-1}:1^{n-\ttk-1}]$}
  \For{$\ttb^\ttk\in[0^{\ttk}:1^{\ttk}]$}
  \State $\ttu[\lsb(\ttk+1)][\ttc^{n-\ttk-1}\ttb^{\ttk}0]
  = \ttu[\lsb(\ttk)][\ttc^{n-\ttk-1}0\ttb^{\ttk}]
  \oplus\ttu[\lsb(\ttk)][\ttc^{n-\ttk-1}1\ttb^{\ttk}]$
  \State $\ttu[\lsb(\ttk+1)][\ttc^{n-\ttk-1}\ttb^{\ttk}1]
  = \ttu[\lsb(\ttk)][\ttc^{n-\ttk-1}1\ttb^{\ttk}]$
  \EndFor
  \EndFor
  \EndFor
 \end{algorithmic}
\end{algorithm}

\subsection{Algorithm for Systematic Channel Encoder}
\label{sec:systematic}

For completeness, we introduce an algorithm for the systematic encoder
of a polar channel code based on \cite{A11}.

The following algorithm finds $x_{\I'_1}$ from $x_{\I'_0}$ and $u_{\I_1}$
such that there is $u_{\I_0}$ satisfying
$(x_{\I'_1},x_{\I'_0})= (u_{\I_1},u_{\I_0}) G$,
where the elements in $(x_{\I'_1},x_{\I'_0})$ and 
$(u_{\I_1},u_{\I_0})$ are sorted in index order before operating $G$.
It should be noted that $(x_{\I'_1},x_{\I'_0})= (u_{\I_1},u_{\I_0}) G$
is equivalent to $(x_{\I'_1},x_{\I'_0}) G = (u_{\I_1},u_{\I_0})$.
We assume that the following algorithms
have access to $\I_1$ and memory spaces
\begin{align*}
 &\lrb{
  \ttX[b^n]:
  b^n\in[0^n:1^n]
 }
 \\
 &\lrb{
  \ttV[b^n]:
  b^n\in[0^n:1^n]
 }.
\end{align*}
The result is stored in $\{\ttX[b^n]\}_{b^n\in[0^n:1^n]}$,
which is also used to refer to the value $u_{\I_1}$
in Algorithm \ref{alg:calcV}.
For a given $b^n\equiv(b_0,\ldots,b_{n-1})$, let $\br(b^n)$ be defined as
\begin{equation*}
 \br(b^n)\equiv (b_{n-1},\ldots,b_0).
\end{equation*}

\begin{algorithm}[H]
 \caption{Systematic Channel Encoder}
 \hspace*{\algorithmicindent}\textbf{Input:}
 $\I_1$, $x_{\I'_0}$, $u_{\I_1}$
 \begin{algorithmic}[1]
  \For{$\ttb^n\in[0^n:1^n]$}
  \If{$\ttb^n\in\I_1$}
  \State $\ttX[\ttb^n]\leftarrow u_{\ttb^n}$
  \Else
  \State $\ttV[\ttb^n]\leftarrow x_{\br(\ttb^n)}$
  \EndIf
  \EndFor
  \State $\calcV(n,\ttb^0)$
  \FFor{$\ttb^n\in[0^n:1^n]$}
  $\ttX[\br(\ttb^n)]=\ttV[\ttb^n]$
 \end{algorithmic}
\end{algorithm}

\begin{algorithm}[H]
 \caption{$\calcV(\ttk,\ttb^{n-\ttk})$}
 \label{alg:calcV}
 \begin{algorithmic}[1]
  \If{$\ttk=0$}
  \IIf{$\ttb^n\in\I_1$} $\ttV[\ttb^n]\leftarrow \ttX[\ttb^n]$
  \Else
  \State $\calcV(\ttk-1,b^{n-{\ttk}}1)$
  \For{$\ttc^{\ttk-1}\in[0^{\ttk-1}:1^{\ttk-1}]$}
  \IIf{$\ttb^{n-\ttk}0\ttc^{\ttk-1}\notin\I_1$}
  $\ttV[\ttb^{n-\ttk}0\ttc^{\ttk-1}]
  \leftarrow
  \ttV[\ttb^{n-\ttk}0\ttc^{\ttk-1}]\oplus\ttV[\ttb^{n-\ttk}1\ttc^{\ttk-1}]$
  \EndFor
  \State $\calcV(\ttk-1,b^{n-{\ttk}}0)$
  \FFor{$\ttc^{\ttk-1}\in[0^{\ttk-1}:1^{\ttk-1}]$}
  $\ttV[\ttb^{n-\ttk}0\ttc^{\ttk-1}]
  \leftarrow
  \ttV[\ttb^{n-\ttk}0\ttc^{\ttk-1}]\oplus\ttV[\ttb^{n-\ttk}1\ttc^{\ttk-1}]$
  \EndIf
 \end{algorithmic}
\end{algorithm}

\subsection{Improvement of Algorithm \ref{alg:updateT}
 by Assuming $\ttT[k][c^{n-k}]\in\{-1,0,1\}$}
\label{sec:updateT}

We introduce an improvement for Algorithm \ref{alg:updateT}
by assuming $\ttT[k][c^{n-k}]\in\{-1,0,1\}$.

For simplicity,
we assume that $\ttT[k][c^{n-k}]$ is represented by a $3$-bit signed integer
consisting of a sign bit and two bits representing an absolute value.

\setcounter{algorithm}{1}
\renewcommand{\thealgorithm}{\arabic{algorithm}'}
\begin{algorithm}[H]
 \caption{$\updateT(\ttT,\ttU,\ttk,\ttb^{\ttk})$}
 \begin{algorithmic}[1]
  \IIf{$k=0$}
  \Return
  \IIf{$\ttb_{\ttk-1}=0$}
  $\updateT(\ttT,\ttU,\ttk-1,\ttb^{\ttk-1})$
  \For{$\ttc^{n-k}\in[0^{n-\ttk}:1^{n-\ttk}]$}
  \If{$\ttb_{\ttk-1}=0$}
  \State $\ttT[\ttk][\ttc^{n-\ttk}]
  \leftarrow
  \ttT[\ttk-1][\ttc^{n-\ttk}1] \cdot \ttT[\ttk-1][\ttc^{n-\ttk}0]$
  \Else
  \State $\ttT[\ttk][\ttc^{n-\ttk}]
  \leftarrow 
  \ttT[\ttk-1][\ttc^{n-\ttk}1]\mp_{\ttu}\ttT[\ttk-1][\ttc^{n-\ttk}0]$,
  where $\ttu\equiv\ttU[\ttk][\ttc^{n-\ttk}][0]$
  \IIf{$\ttT[\ttk][\ttc^{n-\ttk}]<0$}
  $\ttT[\ttk][\ttc^{n-\ttk}]\leftarrow -1$
  \IIf{$\ttT[\ttk][\ttc^{n-\ttk}]>0$}
  $\ttT[\ttk][\ttc^{n-\ttk}]\leftarrow 1$
  \EndIf
  \EndFor
 \end{algorithmic}
\end{algorithm}

\setcounter{algorithm}{12}
\renewcommand{\thealgorithm}{\arabic{algorithm}}

\subsection{Algorithm for Line 5 of Algorithm \ref{alg:prunePath}}
\label{sec:pruePath}

We can implement Line 5 of Algorithm \ref{alg:prunePath} 
by $\markPath(\ttL)$ defined as Algorithm \ref{alg:markPath}.

We assume that Algorithms \ref{alg:markPath}--\ref{alg:swapIndex}
can access the memory space
$\{\ttP[\lambda]\}_{\lambda=0}^{2L-1}$,
$\{\Index[\lambda]\}_{\lambda=0}^{2L-1}$,
and $\{\Active[\lambda]\}_{\lambda=0}^{2L-1}$,
where $\Index[\lambda]\in\{0,\ldots 2L-1\}$ is an integer variable.
The result is stored in $\{\Active[\lambda]\}_{\lambda=0}^{2\ttL-1}$.

\begin{algorithm}[H]
 \caption{$\markPath(\ttL)$}
 \label{alg:markPath}
 \begin{algorithmic}[1]
  \For{$\ttl\in\{0,\ldots,\ttL-1\}$}
  \State $\Index[\ttl]=\ttl$
  \State $\Active[\ttl]=0$
  \EndFor
  \State $\selectPath(0,2\cdot\ttL-1)$
  \FFor{$\lambda\in\{0,\ldots,L-1\}$}
  $\Active[\Index[\ttl]]=1$
 \end{algorithmic}
\end{algorithm}
\begin{algorithm}
 \caption{$\selectPath(\ttleft,\ttright)$}
 \label{alg:selectPath}
 \begin{algorithmic}[1]
  \If{$\ttleft<\ttright$}
  \State $\ttl\leftarrow \partition(\ttleft,\ttright)$
  \IIf{$\ttl>L$}
  $\selectPath(\ttleft,\ttl-1)$
  \IIf{$\ttl<L$}
  $\selectPath(\ttl+1,\ttright)$
  \EndIf
 \end{algorithmic}
\end{algorithm}

\begin{algorithm}[H]
 \caption{$\partition(\ttleft,\ttright)$}
 \label{alg:partition}
 \begin{algorithmic}[1]
  \State Let $\ttl$ be one of the values in $\{\ttleft,\ldots,\ttright\}$
  selected uniformly at random  
  and call $\swapIndex(\ttl,\ttright)$.
  \State $\ttp\leftarrow\ttP[\ttright]$
  \State $\ttl\leftarrow\ttleft-1$
  \State $\ttl'\leftarrow\ttright$
  \Loop
  \State $\ttl\leftarrow \ttl+1$
  \State $\ttl'\leftarrow \ttl'-1$
  \WWhile{$\ttP[\Index[\ttl]]\geq\ttp$ and $\ttl\leq \ttl'$}
  $\ttl\leftarrow \ttl+1$
  \WWhile{$\ttP[\Index[\ttl']]<\ttp$ and $\ttl\leq \ttl'$}
  $\ttl'\leftarrow \ttl'-1$
  \IIf{$\ttl\geq \ttl'$}
  $\textbf{break}$
  \State $\swapIndex(\ttl,\ttl')$.
  \EndLoop
  \State $\swapIndex(\ttl,\ttright)$
  \State
  \Return $\ttl$
 \end{algorithmic}
\end{algorithm}

\begin{algorithm}[H]
 \caption{$\swapIndex(\ttl,\ttl')$}
 \label{alg:swapIndex}
 \begin{algorithmic}[1]
  \State $\tti\leftarrow \Index[\ttl]$
  \State $\Index[\ttl]\leftarrow\Index[\ttl']$
  \State $\Index[\ttl']\leftarrow \tti$
 \end{algorithmic}
\end{algorithm}

\begin{rem}
As mentioned in \cite{TV15},
we can simply sort $\{\Index[\lambda]\}_{\lambda=0}^{\ttL-1}$
so that $\ttP[\Index[0]]\geq\ttP[\Index[1]]\geq\cdots\geq\ttP[\Index[\ttL-1]]$
instead of calling $\selectPath(0,2\cdot\ttL-1)$
at Line 5 of Algorithm \ref{alg:markPath}.
Although the time complexity of sorting is $O(\ttL\log\ttL)$,
it could be faster than $\selectPath(0,2\cdot\ttL-1)$ when $\ttL$ is small.
\end{rem}

\begin{rem}
Line 1 of Algorithm \ref{alg:partition}, which can be omitted,
guarantees that the average time complexity of $\selectPath(0,2\cdot\ttL-1)$
is $O(\ttL)$.
We can replace this line by selecting the index
corresponding to the median of $\{\ttP[\lambda]\}_{\lambda=\ttleft}^{\ttright}$
to guarantee worst-case time complexity $O(\ttL)$ (see \cite{BFPRT73}).
\end{rem}

\end{document}